\documentclass{ifacconf}  

\usepackage{natbib}
\usepackage{epsfig} 
\usepackage{amsmath,amssymb,amsfonts} 
\usepackage{subfigure}
\usepackage{float, color}
\usepackage{enumitem}

\usepackage{theorem}
\theoremstyle{plain} \theorembodyfont{\upshape}
\newtheorem{theorem}{Theorem}
\newtheorem{lemma}{Lemma}
\newtheorem{myassumption}{Assumption}
\newtheorem{myremark}{Remark}
\newtheorem{proposition}{Proposition}
\newtheorem{problem}{Problem}

\def\qed{\hfill $\Box$}
\DeclareMathOperator{\diag}{diag}
\usepackage{natbib}  

\allowdisplaybreaks[1]

\begin{document}

\thispagestyle{empty}
\pagestyle{plain}

\begin{frontmatter}

\title{
A State Feedback Controller for Mitigation of Continuous-Time Networked SIS Epidemics}

\thanks[footnoteinfo]{%
This work was supported in part by  the Knut and Alice Wallenberg Foundation, Wallenberg Scholar under Grant 66469; by a  Distinguished Professor Grant from the Swedish Research Council (Org: JRL, project no: 3058); and by JSPS under Grant-in-Aid for Scientific Research Grant No. 22H01508; {and by the National Science Foundation under Grants No. 2211815 and No. 2213568}}

\author[First]{Yuan Wang}
\author[Second]{Sebin Gracy}
\author[Second]{C\'esar A. Uribe}
\author[Third]{Hideaki Ishii}
\author[First]{and Karl Henrik Johansson}

\address[First]{Division of Decision and Control Systems, School of Electrical Engineering and Computer Science, KTH Royal Institute of Technology, and Digital Futures, Stockholm, Sweden. yuanwang@kth.se,  kallej@kth.se}
 \address[Second]{Department of Electrical and Computer Engineering, Rice University, Houston, TX, USA. sebin.gracy@rice.edu, cauribe@rice.edu}
\address[Third]{Department of Computer Science, Tokyo Institute of Technology, Yokohama, Japan. ishii@c.titech.ac.jp}


\begin{abstract}
The paper considers continuous-time networked susceptible-infected-susceptible (SIS) diseases spreading over a population. Each agent represents a sub-population and {has its own healing rate and infection rate}; the state of the agent at a time instant denotes what fraction of the said sub-population is infected with the disease at the said time instant. By taking account of the changes in behaviors of the agents in response to the
infection rates in real-time, 
our goal is to devise a feedback strategy such that the infection level for each agent strictly stays below a pre-specified value. Furthermore, we are also interested in ensuring that the closed-loop system converges either to the disease-free equilibrium or, when it exists, to the endemic equilibrium. The upshot of devising such a strategy is that it allows health administration officials to ensure that there is sufficient capacity in the healthcare system to treat the most severe cases. We demonstrate the effectiveness of our controller via numerical examples.
\end{abstract}
\end{frontmatter}

\section{Introduction}
\vspace{-1ex}
Mathematical epidemiology has a rich history, tracing its roots to the seminal work by Daniel Bernoulli in \citep{bernoulli1760essai}. Subsequent decades and centuries have witnessed tremendous advancements in this area, with contributions spanning across various  fields such as physics \citep{van2008virus}, computer science \citep{wang2003epidemic}, mathematics \citep{hethcote2000mathematics}, and so on. Some of the central questions pertain to the ascertainment of conditions leading to a disease becoming extinct or persisting in the population. To this end, several models abound in the literature. This paper is focused on the susceptible-infected-susceptible (SIS) model.

The conventional SIS model in \citep{kermack1927contribution} analyzes the spread 
of epidemics from a macroscopic perspective. 
Recently, to study virus spread across population subgroups,  networked SIS models have been proposed for both continuous-time systems  \citep{khanafer2016stability} and discrete-time systems \citep{pare2018analysis}. In this paper, the focus is on \emph{continuous-time networked SIS} models. In this model, an agent is either in the healthy or in the infected state. A healthy agent could get infected depending on its infection rate $\beta$, scaled by the interactions it has with its neighbors; an infected agent recovers based on its healing rate $\gamma$. The terms $\beta$ and $\gamma$ denote the (reciprocal of) time that it takes for an agent to get infected, and to get recovered, respectively.
We say that the networked SIS model is in the healthy state if all the agents are healthy, or equivalently, in the disease-free equilibrium (DFE). If the epidemic remains persistent, we say that the networked SIS model is in the endemic state.

Stability analysis of continuous-time networked SIS models has been studied in the literature \citep{liu2019analysis,ye2021applications,fall2007epidemiological}.
In the context, the results in, among others, \citep{liu2019analysis,ye2021applications,fall2007epidemiological} provide a foundational understanding on what causes an epidemic to become extinct or to become persistent in the population. In other words, the results in these works cater to the question: Assuming that  there are no interventions (pharmaceutical or otherwise), under what conditions would one expect the number of infected {individuals} in a population to grow exponentially? However, one of the main challenges that public health officials confront is ensuring that the available healthcare facilities do not get overwhelmed. In this context, this means that one of the questions of interest is the following: What kind of interventions are needed to guarantee that the proportion of infected in a district or city stays within an acceptable limit?
{One approach towards addressing this question would be to  make the individuals aware of {the proportion of infected} in the population, so that they can
regulate their interactions with others and increase their
means of protection in real-time.} 
From a systems theory viewpoint, the interest, then, is in designing a (possibly closed-loop) control strategy so that the state trajectories ({which means the infected proportion dynamics of each population}) never exceed a certain pre-specified bound. {In this paper, each agent is equipped with a local pre-specified infection level which can be different from agent to agent. From the practical viewpoint, such pre-specified bound is considered as the overwhelmed limit for the healthcare facilities in each agent.  }

Control strategies for epidemic models have been presented in  \citep{pasqualetti2014controllability, preciado2014optimal,liu2019analysis,ye2021applications, morris2020optimal,wang2021suppressing}. More specifically, the control schemes proposed in \citep{pasqualetti2014controllability} are for a class of linearized SIS models. 
The 
paper \citep{preciado2014optimal} studies the problem of cost-optimal distribution of resources (such as vaccines and antidotes) in networked SIS models, whereas, by regarding the healing rate as the local control input, distributed feedback controllers are proposed 
in \citep{liu2019analysis,ye2021applications}. Note that the goals in \citep{preciado2014optimal,ye2021applications,liu2019analysis} do not necessarily ensure the infection levels always staying below a pre-specified level. 
In the absence of drugs and vaccines, the control scheme in \citep{morris2020optimal} regards the effective infection rate of each agent as the control input, and proposed an optimal strategy to reduce the peak in the macroscopic {susceptible-infected-recovered (SIR)} model. A similar control scheme was adopted in \citep{wang2022a} in the context of opinion dynamics.

For discrete-time networked SIS models, a state feedback controller is proposed in \citep{wang2021suppressing}, to guarantee that the fraction of infected in each sub-population stays below half for all time instants. {Note that the discrete-time
SIS model is an approximation 
of the continuous-time SIS model \citep{pare2018analysis}, which in turn is a mean field approximation of a $2^{n}$ state Markov chain model. In order to ensure that the approximation error is low, several additional assumptions are needed on the 
sampling period for the model in \citep{wang2021suppressing}, and furthermore the pre-specified level is the same for all agents. It is natural to ask how well would the controller behave in a continuous-time setting. For this reason, we consider a continuous-time counterpart of the work \citep{wang2021suppressing} in this paper. Furthermore, we  do not insist on the pre-specified level being the same for all agents, and we are interested in guaranteeing that even for the closed-loop SIS system,  {the local proportion of infected individuals} in each node  converges either to zero or to a  scalar that is strictly less than said pre-specified level.}

\emph{Contributions:} 
The main contribution of this paper is to devise a control scheme for guaranteeing that the fraction of  infected individuals in a sub-population stays within a pre-specified level for all time instants. 
Our approach is as follows: First, we modify the continuous-time SIS model in \citep{liu2019analysis} by introducing a parameter that scale the strength of interconnections between agents, based on their current infection levels.  We then show that for this modified SIS model the following properties hold:
\begin{enumerate}[label=(\roman*)]
    \item The proportion of infected {individuals} in a sub-population does not exceed a pre-specified level; see Proposition~\ref{lemma0}.
    \item If  the spectral abscissa of {the state matrix linearized around the healthy state} is not greater than zero, then the DFE is asymptotically stable; see Proposition~\ref{theorem1a}
    \item If  the spectral abscissa of {the state matrix linearized around the healthy state}  is greater than zero, then there exists an endemic equilibrium to which the dynamics converge for all non-zero initial conditions; see Proposition~\ref{theorem6}.
\end{enumerate}

This paper is structured as follows. In Section~\ref{sec:2}, the continuous-time networked SIS model and feedback control problem of interest are formally introduced. Our main results and related discussions are given in Section~\ref{sec:3}. A numerical example is presented to illustrate our theoretical findings in in Section~\ref{sec:4}. Finally, we conclude this paper, and provide future directions in Section~\ref{sec:5}. 

\emph{Notation:}
Let $\mathbb{R}_+$ denote the sets of non-negative real numbers. We use $[n]$ to denote the set $\{1,2,...,n\}$ for any positive integer $n$. For any two vectors $\mathbf{a},\mathbf{b} \in \mathbb{R}^{n}$, we write {$\mathbf{a} > \mathbf{b}$} if $a_i > b_i$ for every $i \in [n]$. For a matrix $A$, let $s(A)$ denote the largest real part among the eigenvalues of $A$. A diagonal matrix is denoted as $\diag(\cdot)$. 



\section{Problem Formulation}
\label{sec:2}
\vspace{-1ex}
\subsection{Open-loop networked SIS model}


Consider a disease spreading over a network of $n$ agents. The interconnection among the various agents is represented by a directed graph $\mathcal{G}=(\mathcal{V},\mathcal{E})$, where $\mathcal V =\{1,2,\ldots,n\}$ is the set of agents. Let $A=[a_{ij}]_{n \times n}$ be the weighted adjacency matrix of $\mathcal G$. The edge set $\mathcal E$ is defined as follows: $\mathcal E =\{(j,i) \mid a_{ij}>0\}$. More specifically, if a pair of vertices $(j,i) \in \mathcal{E}$, there is a directed edge from node $j$ to node $i$. The set of in-neighbors for agent $i$ is denoted as $\mathcal{N}_i$.   Throughout the rest of this paper, an agent can be interpreted as a \emph{sub-population}. The weight between agents can be considered as the frequency of interaction between the sub-populations. Each sub-population is comprised of individuals and the number of individuals in a sub-population is fixed. Each individual is assumed to be either Infected (I) with the disease, or not infected but Susceptible (S) to the disease.
The disease can spread both due to the interaction between individuals in a sub-population and also across sub-populations.


Let $x_i(t) \in [0,1]$ denote the proportion of (sub)population $i \in [n]$ 
that is infected with the disease at time $t \in \mathbb{R}_+$ (thus $1-x_i(t)$ is the proportion of (sub)population $i$ who are susceptible). Let $\beta_i$ and $\gamma_i$ denote the infection rate and healing rate, respectively, of agent $i$.
The evolution of the infection level of agent $i$ can, then, be 
captured by the following differential equation~\citep{fall2007epidemiological}:
\begin{equation} \label{eq:ct}
 \dot{x}_{i}(t) = \beta_i (1-x_{i}(t))  \textstyle\sum_{j=1}^{n}a_{ij}x_{j}(t)- \gamma_i x_{i}(t), 
\end{equation}
where $\beta_i > 0$ and $\gamma_i \geq 0$. 

Define $\mathbf{x}(t) := [x_1(t), x_2(t), \ldots, x_n(t)]^T$, $ B := \mathrm{\diag}(\beta_i)$, $\Gamma := \mathrm{\diag}(\gamma_i)$, 
and $X(t): =\mathrm{\diag}(\mathbf{x}(t))$ being the diagonal matrices. Then in vector form the open-loop system~\eqref{eq:ct} can be written as
\begin{equation} \label{eq:ctv}
\mathbf{\dot{x}}(t) =   \big[(I-X(t))BA
                        -  \Gamma \big]\mathbf{x}(t).
\end{equation}
It is clear that $\mathbf{0}_n$ 
is an equilibrium of the open-loop system~\eqref{eq:ctv}. This equilibrium point is referred to as \textit{the healthy state}, or the \textit{disease-free equilibrium} (DFE).
Any equilibrium other than $\mathbf{0}_n$ is 
referred to as the endemic equilibrium, under which the disease persists in at least one agent.

We need the following assumptions to ensure that system~\eqref{eq:ctv} is well-defined.
\begin{myassumption} \label{assumption:pos:syspar}
For each $i \in [n]$, $\beta_i > 0$ and $\gamma_i \geq 0$.
\end{myassumption}

\begin{myassumption} \label{assumption1}
The matrix $A$ is irreducible.
\end{myassumption}

Assumption~\ref{assumption:pos:syspar} stipulates 
that all infection and healing rates must be positive and non-negtaive, respectively.
Assumption~\ref{assumption1} states 
that the adjacency matrix $A$ is irreducible, which is fulfilled if and only if the underlying graph $\mathcal{G}$ is strongly connected. A graph $\mathcal G$ is strongly connected if and only if each node $i \in \mathcal V$ has a path to every other node $j \in \mathcal V$.

\subsection{Modified SIS models with local control inputs}

Our 
objective is to ensure that, for all $i \in [n]$ and $t>0$, $x_i(t)$ remains upper bounded by a constant ${1}/{c_i}$, for some $c_i>1$, and that the dynamics converge to one of the equilibria of the closed-loop system.  That is, we are interested in ensuring that the fraction of infected population in each node stays below a pre-specified value for all times.  Furthermore, we would like to ensure that $\lim_{t \rightarrow \infty}x_i(t)=\bar{x}_i$, where $\bar{x}_i$ denotes some equilibrium of the closed-loop system. 
To this end, each agent $i \in [n]$ is equipped with a local control input $u_i(t)$. As a consequence, System~\eqref{eq:ct} can be modified as follows: 
\begin{align} \label{eq:dt+control}
 \dot{x}_{i}(t) =  \beta_i(1-x_{i}(t)) \textstyle\sum_{j=1}^{n}a_{ij}x_{j}(t) - \gamma_i x_{i}(t) + u_i(t),
\end{align}
where $u_i(t)$  is a state feedback control of the form:
\begin{align}\label{eq:u0}
u_i(t)=f_i\left(x_i(t), \{ x_j(t)\}_{j\in \mathcal{N}_i}\right).
\end{align}
{The state feedback controller $u_i(t)$ can be considered as the expected reduction for growth rate of infection at time $t$ because of pharmaceutical or non-pharmaceutical interventions. 
From a practical standpoint, 
the non-pharmaceutical interventions may include hand-washing, mask-wearing, social distancing, etc. The typical pharmaceutical intervention is the distribution of immunity-boosting or therapeutic drugs. A rough approach to estimating such $u_i(t)$ is to compare the growth rate of infection before and after the interventions in a short time period. 
However, we note that these consequences of the mentioned pharmaceutical or non-pharmaceutical interventions are very hard to measure or estimate accurately.  
There are some practical limitations on how the mentioned interventions can affect the spread. The effective disease prevention measures, according to the type of disease, are quite different. As a simple example, mask-wearing may slow down the spreading of airborne diseases. However, it may not mitigate the infection of waterborne diseases. The modeling for effects of detailed interventions are out of scope in this paper.}
{See Remarks~\ref{prac:inter} and~\ref{remark2} for a slightly more detailed explanation for the proposed feedback controller.} 
\par We formally state the problem being investigated in this paper.

\begin{problem}\label{q1}
Consider the closed-loop System~\eqref{eq:dt+control}. Based on knowledge of $\beta_i$, $x_i(t)$, $c_i$, $a_{ij}$ and $x_j(t)$ where $j \in \mathcal{N}_i$, 
design a local state feedback controller in accordance with~\eqref{eq:u0} such that
\begin{enumerate}[label=\roman*)]
    \item for all $i \in [n]$ and $t > 0$, $x_i(t) \leq {1}/{c_i}$; and
    \item $\lim_{t \rightarrow \infty} x_{i}(t) =x_i^*$, where either $x_i^*=0$ or $x_i^*>0$.
\end{enumerate}
\end{problem}
A discrete-time version of Problem~\ref{q1} has been partially solved in \citep{wang2021suppressing} 
with the local state feedback controller having a specific structure, {namely, i) $c_i=2$ for all $i \in [n]$, and ii) the strength of interconnections modified as a function of the current infection levels of each node and all of its neighbors}.

Our approach towards addressing Problem~\ref{q1} involves two steps: First,  given some $c_i>1$, we identify $u_i(t)$ that obeys~\eqref{eq:u0}, such that $x_i(t) \leq {1}/{c_i}$ for all $t$. Second, we show that, for the proposed feedback controller $u_i(t)$, the closed-loop system~\eqref{eq:ct} always converges to some equilibria. 
To elucidate more on the second part of our solution, we now recall a result which fully characterizes the class of equilibria for the open-loop system~\eqref{eq:ct}, and admits $\gamma_i = 0$.
\begin{proposition}\cite[Propositions~2 and 3]{liu2019analysis}
\label{prop0}
Consider the open-loop System \eqref{eq:ctv} 
{
under Assumptions~\ref{assumption:pos:syspar} and~\ref{assumption1}.}
The following statements hold:
\begin{enumerate}
\item If $s(BA-\Gamma) \leq 0$, then, the disease-free equilibrium is the unique equilibrium, and it is globally asymptotically stable.
\item If $s(BA-\Gamma) > 0$, then, other than the disease-free equilibrium, there exists a unique endemic equilibrium, which is globally asymptotically stable on $[0,1]^n \setminus \{\mathbf{0}_n\}$.
\end{enumerate}
\end{proposition}

In the rest of this paper, we will
show that the result in Proposition~\ref{prop0} can be, with a suitable choice of $u_i(t)$ as in~\eqref{eq:u0}, extended for the closed-loop system in~\eqref{eq:dt+control}.
\section{Main results}
\label{sec:3}
\vspace{-1ex}
In this section, we devise a 
local state feedback controller to solve Problem~\ref{q1}.
{Observe that 
for the closed-loop System~\eqref{eq:dt+control}, since the term $\beta_i(1-x_{i}(t)) \textstyle\sum_{j=1}^{n}a_{ij}x_{j}(t)$ is positive, the terms that contribute to the increase in the state value are exactly the aforementioned ones. Hence, in order to ensure that the state trajectories remain below $1/c_i$ for all times, it is natural to consider the controller that partially offsets the increase caused by the aforementioned terms. For this reason,}
we consider the following control input 
\begin{align}\label{eq:u}
  u_i(t) = - \beta_i c_ix_i(t)(1-x_i(t)) \textstyle\sum_{j=1}^{n} a_{ij}x_j(t).  
\end{align}
Consequently, by plugging~\eqref{eq:u} into~\eqref{eq:dt+control}, we obtain:
\begin{flalign}\label{eq-04}
   \dot{x}_i(t) = \beta_i(1-c_ix_i(t))(1-x_i(t)) \textstyle\sum_{j=1}^{n} a_{ij}x_j(t) - \gamma_i x_i(k). 
\end{flalign}

Let $D :=\mathrm{\diag}(c_i)$. Then, the closed-loop System \eqref{eq-04} can be written as 
\begin{flalign} \label{eq-05}
    \mathbf{\dot{x}}(t) =   \big[ (I-DX(t))(I-X(t))BA
                        -  \Gamma \big]\mathbf{x}(t). 
\end{flalign}
We refer to System~\eqref{eq-05} 
as the \emph{controlled system}.

\begin{myremark}\label{prac:inter}
Define $b_i(t): = 1-c_ix_i(t)$, and observe that $b_i(t) \in [0,1]$. That is, the term $b_i(t)$ 
scales the infection rate for agent $i$ at time instant~$t$. More concretely, and especially from a practical standpoint, the parameter $b_i(t)$ may be thought of as a term/quantity, possibly communicated to all the agents by a central entity such as health administration officials, that informs each agent as to by how much should it modify the strength of its interactions with its neighbors.
As a consequence, the infection rate of agent $i$ is reduced  from $\beta_i$ to $b_i(t)\beta_i$. 
\end{myremark}

{
\begin{myremark}\label{remark2}
The proposed feedback controller is based on the ideal environment, and there are certain assumptions made for practical implementation as follows: Agents must be able to get the exact states, including their own and their neighbors, and the proposed control strategy must be faithfully implemented by agents. We must implement that, in practice, there are asymptomatic members (who are in the latent period), and it is hard to handle the exact proportion of infected people. The design of a feedback controller which is based on noisy state estimation, as well as the epidemic control policy within the 
antagonistic environment (where some agents may not follow the control strategy) are left for our future studies.
\end{myremark}
}

We have the following assumption.

\begin{myassumption} \label{assumption2}
For each $i \in  [n]$, $x_i(0) \in [0, 1/c_i]$.
\end{myassumption}

Assumption~\ref{assumption2} ensures that all initial states are within the desired bound, since otherwise the control objectives cannot be achieved. 

With $u_i(t)$ as in~\eqref{eq:u0}, the infection levels in the closed-loop System~\eqref{eq:dt+control} do not exceed the pre-specified level, ${1}/{c_i}$.
We have the following propositions. {We skip the proof of the propositions due to space reasons.}
\begin{proposition}\label{lemma0}
Consider the controlled System \eqref{eq-05} under Assumptions~\ref{assumption:pos:syspar} and~\ref{assumption2}. 
Then $x_i(t) \in [0,1/c_i]$ for all $i \in [n]$ and $t \geq 0$.
\end{proposition}

Proposition~\ref{lemma0} establishes that the set $[0,1/c_i]$ is positively invariant for each agent $i$ with respect to the closed-loop System~\eqref{eq-04}. Positive invariance of the set $[0,1]^n$ with respect to the uncontrolled System~\eqref{eq:ct} has been established in \citep[Lemma~8]{liu2019analysis}. Under Assumptions~\ref{assumption:pos:syspar} and~\ref{assumption2}, the proposed controller 
shrinks the positive invariant set 
for each agent $i$ from $[0,1]$ to $[0,1/c_i]$. In a practical sense, the local infection upper bound $1/c_i$ can be regarded as the capacity of healthcare facilities in a local population $i$. As one of the main advantages of the controlled system, if the initial infection level do not get overwhelmed, then the epidemic remains manageable.

While Proposition~\ref{lemma0} guarantees that the infection level does not exceed a certain pre-specified level, it does not give any guarantees regarding the long-term behavior of the system. That is, with the aforementioned controller in place, does the disease die out, or does it become endemic? We address the same in the rest of this subsection. 

It turns out that convergence guarantees, similar to that for the open-loop case discussed in Proposition~\ref{prop0},  can be given even for the closed-loop System~\eqref{eq-05}. The next two propositions handle the case when 
$s(BA-\Gamma) \le 0$ and $s(BA-\Gamma) > 0$, respectively.

\begin{proposition} \label{theorem1a}
Consider the controlled System~\eqref{eq-05} under Assumptions \ref{assumption:pos:syspar}--\ref{assumption2}.
If $s(BA-\Gamma) \le 0$, then the disease-free equilibrium is asymptotically stable with the domain of attraction $[0,1/c_i]$ for each $i \in [n]$.  
\end{proposition} 



Proposition~\ref{theorem1a} states that, irrespective of whether an agent is initially infected or healthy, as long as $s(BA-\Gamma) \le 0$ the infection levels of the closed-loop system converge to the disease-free equilibrium. Note that Proposition~\ref{lemma0} depends only on the healing rate, infection rate and the network structure. If the healing rate of each agent dominates the infection rate (which is scaled by the interconnection weights between agents), the epidemic dies out naturally.  
\begin{proposition} \label{theorem6}
Consider the controlled System~\eqref{eq-05} under Assumptions \ref{assumption:pos:syspar}--\ref{assumption2}.
If $s(BA-\Gamma) > 0$, then there exists a unique endemic equilibrium $\overline{\mathbf{x}}$ such that $0< [\overline{\mathbf{x}}]_i \leq {1}/{c_i}$ for each $i \in [n]$, and $c_i>1$.
Furthermore, if for each $i \in [n]$, $c_i\geq 2$, then the endemic equilibrium is asymptotically stable with the domain of attraction $(0,1/c_i]$. 
\end{proposition}

Proposition~\ref{theorem6} states that when $s(BA-\Gamma) > 0$ System~\eqref{eq-05} has two equilibria, namely, the disease-free equilibrium and the endemic equilibrium $\mathbf{\overline{x}}$. Furthermore, as long as the initial state is non-trivial and $c_i \geq 2$, the state of the closed-loop system
asymptotically converges to $\mathbf{\overline{x}}$, which means that the disease-free equilibrium is unstable. 
In other words,
our proposed controller cannot eradicate the disease;
rather, it ensures that each population $i$ has a stable non-trivial infection level strictly smaller than ${1}/{c_i}$.

We are now ready to state the main result of the paper. It summarizes the results from Propositions~\ref{lemma0}--\ref{theorem6}. 
Recall that the overall closed-loop
system in~\eqref{eq-05} consists of the agent dynamics in~\eqref{eq:dt+control} and the local control inputs in~\eqref{eq:u}.

\begin{theorem}\label{thm:prob:1}
Consider the closed-loop System~\eqref{eq:dt+control} under Assumptions~\ref{assumption:pos:syspar}--\ref{assumption2}. The local nonlinear state feedback control law~\eqref{eq:u}, guarantees $x_i(t) \leq {1}/{c_i}$ for all $i \in [n]$ and $t \geq 0$. Furthermore, if $s(BA-\Gamma) \leq 0$, then $\lim_{t \rightarrow \infty} x_i(t) =0$; otherwise if $s(BA-\Gamma) > 0$, and $c_i \geq 2$ for all $i \in [n]$, then $\lim_{t \rightarrow \infty} x_i(t) = x_i^*$, where $x_i^*>0$.
\end{theorem}

{As a practical consequence, Theorem~\ref{thm:prob:1} indicates how the faithfully implemented epidemic prevention interventions influence the infection spreading.}
Note that the convergence result for the case when $s(BA-\Gamma) > 0$, and $1< c_i < 2$ is not provided.
From a technical viewpoint, the main hindrance is that the current Lyapunov candidate is not strictly decreasing for the aforementioned case. 
Simulations, however, indicate that even for the case when $s(BA-\Gamma) > 0$, and $1< c_i < 2$ convergence to the endemic equilibrium can be achieved. A rigorous analysis of the aforementioned case is left for future work.


\section{Numerical Example}
\label{sec:4}
In this section, we first illustrate the results from Section~\ref{sec:3}, and then show some interesting behavior via simulations.

\begin{figure}[t]
  \centering
      \includegraphics[width=0.8\linewidth]{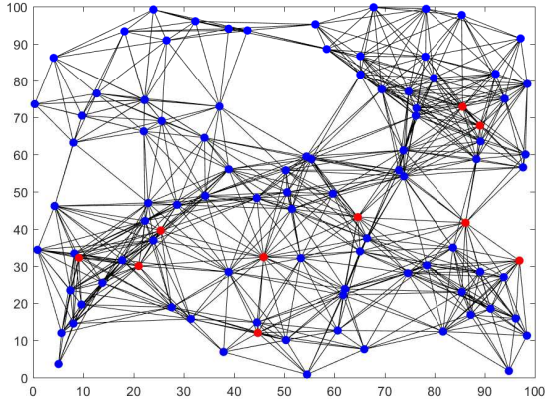}
      \caption{Network topology 1 with communication radius \mbox{$r=25$}.}
      \label{fig1}
\end{figure}

The simulations are presented in a multi-agent network with $n=100$ sub-populations. The sub-populations are uniformly and randomly located in an area of $100 \times 100$. 
All agents within a distance of $r=25$ are allowed to communicate with each other.
{Following a geometric graph model}, for any two sub-population 
$i$ and $j$
with the distance between them being less than $r$, there exists an undirected edge between the said nodes,  with the weight $0<a_{ij}<1$.
The 
nodes
are indexed from $1$ to $100$. Each simulation follows the same initial conditions, such that the 
nodes
$1$ to $10$ have $10$ percent of their population infected, 
and the other 
nodes, i.e., numbered $11$ to $100$ are susceptible. 
The first topology of the multi-agent network is as depicted in Fig.~\ref{fig1}, where the nodes that are healthy (or susceptible) are marked blue, while the nodes that are infected are marked red. 
It can be 
easily verified that this network is strongly connected. The 
edge weights $a_{ii}$ and $a_{ij}$ are  
set to $a_{ii}=0.3, a_{ij}=0.003$, for all connected 
nodes $i$ and $j$ such that $a_{ij} \neq 0$.

The pre-specified infection level $1/c_i$ is chosen as follows: For $1 \le i \le 20$,  choose $1/c_i =0.5$; 
for $21 \le i \le 40$, choose $1/c_i =0.45$;
for $41 \le i \le 60$,  choose $1/c_i =0.3$; 
for $61 \le i \le 80$, choose $1/c_i =0.25$;
and for $81 \le i \le 100$, choose $1/c_i =0.2$. Hence, for all $i \in [n]$, the initial states, which are at most $0.1$, are strictly less than $1/c_i$.

\begin{figure}[t]
  \centering
      \includegraphics[width=1\linewidth]{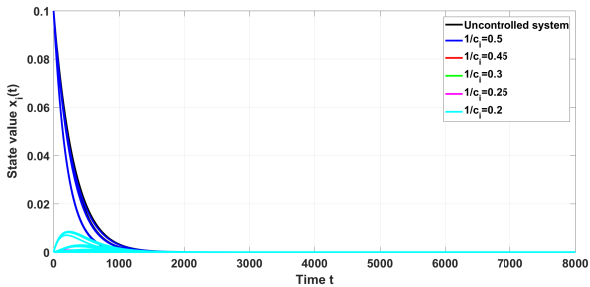}
      \caption{Time responses for both uncontrolled and controlled systems when $s(BA-\Gamma) = -0.313$.}
      \label{fig2}
\end{figure}

We illustrate the case $s(BA-\Gamma) \leq 0$ in the first simulation.
Set $\beta_i = 0.3$, $\gamma_i = 0.5$ for all $i$. Then, we have $s(BA-\Gamma) = -0.313$ for the network in Fig.~\ref{fig1}. The trajectories of the states are shown in Fig.~\ref{fig2}. Both the states in uncontrolled and controlled systems stays strictly below the minimum pre-specific infection level $1/c_i =0.2$. Furthermore, the 
dynamics of the closed-loop system
converges to the disease-free equilibrium, which is in 
line
with the result in Theorem~\ref{thm:prob:1}.

\begin{figure}[t]
  \centering
      \includegraphics[width=1\linewidth]{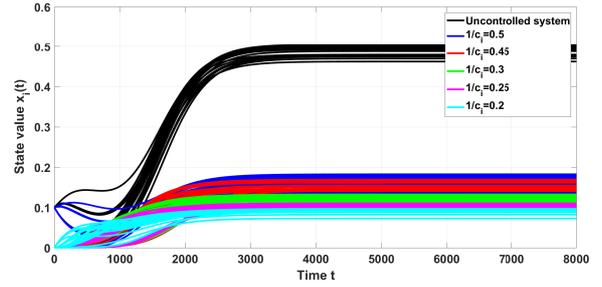}
      \caption{Time responses for both uncontrolled and controlled systems when $s(BA-\Gamma) = 0.198$, and $c_i\geq 2$.}
      \label{fig3}
\end{figure}

Then, we study the case where $s(BA-\Gamma) > 0$. Set $\beta_i = 0.8$, $\gamma_i = 0.3$ for all $i$, and 
consequently, 
$s(BA-\Gamma) = 0.198$.  
The disease persists in both the uncontrolled and controlled systems; see Fig.~\ref{fig3}.
The uncontrolled system converges to the endemic equilibrium around $0.48$. For the controlled system, 
the states asymptotically converge to the endemic equilibrium, and they are all strictly below the minimum pre-specific infection level $1/c_i =0.2$, which is consistent with the result in Theorem~\ref{thm:prob:1}. In the controlled system, even though the disease persists, the infection levels are significantly suppressed in all sub-populations.
Comparing the states with different pre-specific infection levels, the sub-populations that choose a smaller $1/c_i$ generally have a lower infection level.

\begin{figure}[t]
  \centering
      \includegraphics[width=1\linewidth]{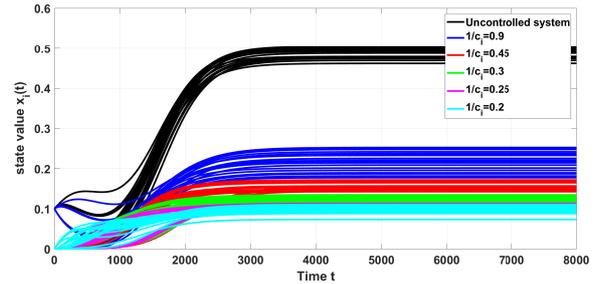}
      \caption{Time responses for both uncontrolled and controlled systems when $s(BA-\Gamma) = 0.198$ and $c_i<2$.}
      \label{fig4}
\end{figure}

We check the convergence result for the case when $s(BA-\Gamma) > 0$, and $1< c_i < 2$. Set $\beta_i = 0.8$, $\gamma_i = 0.3$ for all $i$, and consequently, 
$s(BA-\Gamma) = 0.198$. For $1 \le i \le 20$, we reset the pre-specified infection level by $1/c_i =0.9$, and thus $1<c_i<2$ for these agents. The agent dynamics are shown in Fig.~\ref{fig4}. It can be seen that the infection level for $1 \le i \le 20$ (blue lines) reach the endemic equilibrium.

\begin{figure}[t]
  \centering
      \includegraphics[width=1\linewidth]{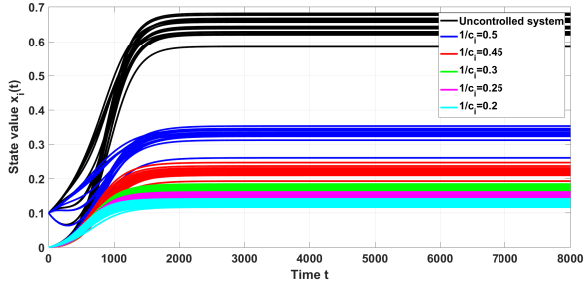}
      \caption{Time responses for both uncontrolled and controlled systems when $s(BA-\Gamma) = 0.276$ and $r =50$.}
      \label{fig5}
\end{figure}

We next study how the higher proximity between agents affects the spreading dynamics. Keep the location of all agent as that in Fig.~\ref{fig1}, and improve the communication radius by $r=50$. Set $\beta_i = 0.8$, $\gamma_i = 0.3$ for all $i$, and consequently, $s(BA-\Gamma) = 0.276$. The time responses are shown in  Fig.~\ref{fig5}. The disease spreading is strictly suppressed below the local pre-specific infection levels in all sub-populations. However, comparing with the time responses in Fig.~\ref{fig3} (where communication radius is $r=25$), the disease persists with higher infection proportions in both the uncontrolled and controlled systems. 

\begin{figure}[t]
  \centering
      \includegraphics[width=0.8\linewidth]{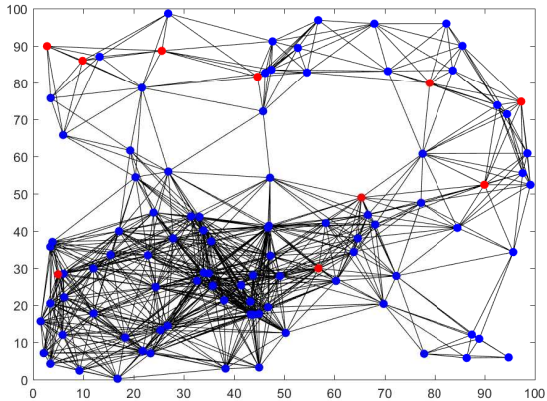}
      \caption{Network topology 2 with communication radius \mbox{$r=25$}.}
      \label{fig6}
\end{figure}

\begin{figure}[t]
  \centering
      \includegraphics[width=1\linewidth]{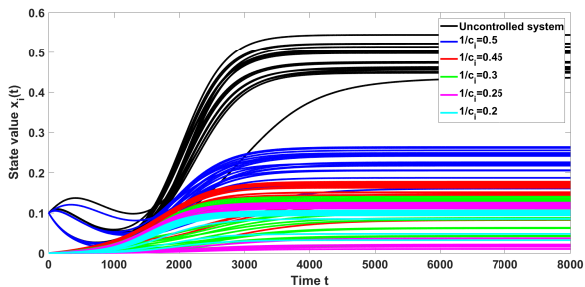}
      \caption{Time responses within network topology 2 for both uncontrolled and controlled systems when $s(BA-\Gamma) = 0.214$ and $c_i\geq 2$.}
      \label{fig7}
\end{figure}

Finally, we check the influence of higher concentration of agents. We test the disease spreading within a network topology as shown in Fig~\ref{fig6}, where the communication radius is set by $r=25$. Comparing with that in Fig~\ref{fig1}, there is a cluster of sub-populations on the left bottom side. Set $a_{ii}=0.3$, $a_{ij}=0.003$, and $\beta_i = 0.8$, $\gamma_i = 0.3$ for all $i$. Consequently, $s(BA-\Gamma) = 0.214$. The infection dynamics are shown in Fig.~\ref{fig7}. Comparing with that in Fig.~\ref{fig3}, it can be seen that with higher concentration of agents, the disease persists with higher infection proportions. As a practical consequence of the simulation results, avoiding crowd aggregation is beneficial to the suppression of infectious diseases.

\section{Conclusion}
\label{sec:5}

We considered an SIS epidemic spreading over a networked population. We designed a state feedback controller where the objective was to ensure that the infection level of each agent never exceeded a pre-specified value. With the controller in place, we identified a condition under which the disease gets eradicated (resp. remains persistent in the population). One of the future directions would be to extend both the feedback controller and analysis approaches for the other disease-spreading models, such as SIR, SEIR, etc. Another interesting direction would be to design state feedback controllers that maintain the infection below a certain level for time-varying graphs, as well as the epidemic control strategy with noisy data in the antagonistic environment.

\bibliography{References-Yuan}

\end{document}